\documentclass[]{aastex701}
\usepackage{float}
\usepackage{amsmath}
\usepackage{booktabs}
\usepackage{textgreek}
\usepackage[utf8]{inputenc}
\usepackage{tikz}
\usetikzlibrary{shapes.geometric, arrows, positioning}
\tikzstyle{startstop} = [rectangle, rounded corners, minimum width=3cm, minimum height=1cm,text centered, draw=black]
\tikzstyle{process} = [rectangle, minimum width=3cm, minimum height=1cm, text centered, draw=black]
\tikzstyle{decision} = [diamond, minimum width=3cm, minimum height=1cm, text centered, draw=black, aspect=2]
\tikzstyle{arrow} = [thick,->,>=stealth]

\begin{document}

\title{High-Speed Observations of Lunar Impact Flashes}

\author[0009-0004-2806-1435]{Dale P. Giancono}
\affiliation{Space Science and Technology Centre, Curtin University, GPO Box U1987, Perth WA 6845, Australia}
\affiliation{International Centre for Radio Astronomy Research, Curtin University, GPO Box U1987, Perth WA 6845, Australia}
\email{dale.giancono@postgrad.curtin.edu.au}

\author[0000-0001-9226-1870]{Hadrien A. R. Devillepoix}
\affiliation{Space Science and Technology Centre, Curtin University, GPO Box U1987, Perth WA 6845, Australia}
\affiliation{International Centre for Radio Astronomy Research, Curtin University, GPO Box U1987, Perth WA 6845, Australia}
\email{Hadrien.Devillepoix@curtin.edu.au}

\author[0000-0002-5864-105X]{Robert M. Howie}
\affiliation{Space Science and Technology Centre, Curtin University, GPO Box U1987, Perth WA 6845, Australia}
\email{robert.howie@curtin.edu.au}

\author[]{Evan Dilley}
\affiliation{University of Western Australia, Perth, WA, Australia}
\email{evan.dilley@uwa.edu.au}

\author[0000-0002-9077-2025]{Bruce Gendre}
\affiliation{University of the Virgin Islands, 2 John Brewer's Bay, St Thomas, 00802,
VI, USA}
\email{bruce.gendre@uwa.edu.au}

\author[0000-0002-7795-9354]{David Coward}
\affiliation{Department of Physics, University of Western Australia, M013, Crawley WA 6009, Australia}
\email{david.coward@uwa.edu.au}

\author[0000-0001-6537-5821]{John Moore}
\affiliation{University of Western Australia, Perth, WA, Australia}
\email{john.moore@uwa.edu.au}

\author[0000-0003-1955-628X]{Sophie E. Deam}
\affiliation{Space Science and Technology Centre, Curtin University, GPO Box U1987, Perth WA 6845, Australia}
\affiliation{International Centre for Radio Astronomy Research, Curtin University, GPO Box U1987, Perth WA 6845, Australia}
\email{sophie.deam@postgrad.curtin.edu.au}

\author[]{Dean Hooper}
\affiliation{International Occultation Timing Association (IOTA), P.O. Box 423, Greenbelt, MD 20768, USA}
\affiliation{Trans-Tasman Occultation Alliance (TTOA), Wellington PO Box 3181, New Zealand} 
\email{dean@hooper.net.au}

\author[0000-0003-1646-6932]{Daniel Sheward}
\affiliation{University of Leicester, School of Physics and Astronomy, University Road, LE1 7RH Leicester, UK}
\email{ds869@leicester.ac.uk}


\begin{abstract} 
Lunar impact flashes provide a direct means of estimating the flux of centimetre-sized meteoroids impacting the lunar surface. However, 25--60 frames per second imaging typical of most monitoring programs limit the ability to resolve the rapid temporal evolution of the impact process, while the integration of Earthshine background restricts the detection of faint flashes. In this work, we present high-speed observations of lunar impact flashes captured at 200 and 250~FPS using the Zadko Telescope in Western Australia. We resolve the light curves of four confirmed events, revealing complex morphologies, some of which are not well modelled by simple exponential decays. One event was simultaneously detected by a second observer using a 50~FPS system, revealing a significantly faster brightness drop in the high-speed data that cannot be explained by spectral differences alone, indicating temporal integration of the vapour plume and subsequent ejecta. Our data also indicates that the initial flash intensity (representing the vapour plume) exhibits significantly less variance across events than the total luminous energy. Furthermore, we found no statistical correlation between the initial luminous energy and the total integrated energy of the flashes in this data, suggesting that the physical mechanism driving the initial vapour expansion may be physically decoupled from the longer-duration glow driven by the cooling ejecta. High temporal resolution combined with high sensitivity are therefore essential for accurately characterising the physical properties of the impactor and distinguishing the initial vapour plume from the subsequent incandescent cooling phase, although a significantly larger dataset is required to definitively constrain these mechanisms. 
\end{abstract}

% \keywords{\uat{Galaxies}{573} --- \uat{Cosmology}{343} --- \uat{High Energy astrophysics}{739} --- \uat{Interstellar medium}{847} --- \uat{Stellar astronomy}{1583} --- \uat{Solar physics}{1476}}


\section{Introduction}

Meteoroids frequently strike the Moon's surface at great velocities, converting their kinetic energy into heat and the excavation of a crater. A small fraction of this energy is released as a brief, transient optical event known as a Lunar Impact Flash (LIF) \citep{ortizOpticalDetectionMeteoroidal2000}. Monitoring these flashes is an important technique for studying the flux and size-frequency distribution of matter in the near-Earth environment, as the Moon acts as a giant, natural detector for meteoroids ranging from centimetres to meters in size \citep{ortizDetectionSporadicImpact2006}. Characterising this meteoroid population in this region is crucial for assessing the impact hazard to artificial satellites and for designing appropriate shielding for future crewed missions and permanent lunar habitats \citep{suggsNASALunarImpact2008}. These natural collisions also offer a unique opportunity to study the physics of hypervelocity impacts at speeds of up to 72 km/s, far exceeding what can be achieved in laboratory experiments \citep{yanagisawaLightcurves1999Leonid2002}. Additionally, precisely locating these impacts in space and time provides known seismic sources that are essential for calibrating future lunar seismometer networks and investigating the Moon's internal structure \citep{lognonneMoonMeteoriticSeismic2009,yamadaOptimisationSeismicNetwork2011}.

The optical flash generated by a lunar impact typically lasts less than 100 milliseconds \citep{liakosNELIOTANewResults2024} and consists of two distinct phases, a short lived highly luminous expansion of a high temperature vapour plume followed by a longer lasting glow attributed to thermal blackbody radiation from an ejected cloud of cooling molten droplets and incandescent particles \citep{yanagisawaLowDispersionSpectra2025}. Numerical simulations of high-velocity impacts predict the initial vapour plume to be an extremely brief event, decaying on a timescale of less than one millisecond \citep{artemievaLunarLeonidMeteors2000}. This is supported by laboratory hypervelocity impact experiments using high-speed cameras and photodetectors, which have consistently measured the rise and decay of the initial light flash on the order of microseconds \citep{tandyImpactFlashEvolution2020,eichhornMeasurementsLightFlash1975,kadonoObservationExpandingVapor1996,ernstEffectInitialConditions2003, ernstHighSpeedImagingImpact2010}. Consequently, standard frame rate observations likely integrate this brief peak luminosity over the full duration of the exposure, potentially leading to an underestimation of the peak flash brightness, peak flash temperature, and a loss of the detailed light curve profile required to distinguish between the different phases of impact evolution \citep{avdellidouTemperaturesLunarImpact2019, avdellidouImpactsMoonAnalysis2021,yanagisawaLowdispersionSpectralVideo2022}. This fundamental mismatch between the physical phenomenon's short duration and the tens-of-milliseconds integration time of standard video cameras (e.g., 20 ms exposures at 50 FPS) also creates an observational bias where faint flashes are averaged out by the background noise. In standard video surveys, the vast majority of flashes appear as single frame events, requiring simultaneous validation by a second observer to distinguish them from cosmic rays, satellites, and electronic noise. \citep{suggsFluxKilogramSizedMeteoroids2014, liakosNELIOTALunarImpact2019}. The correlation between flash magnitude and duration suggests that events as faint as the Earthshine background level persist for only a few milliseconds, thereby remaining undetectable to standard frame rate systems \citep{bouleyPowerDurationImpact2012}. Furthermore, laboratory hypervelocity experiments indicate that the morphology of the light curve is not solely dependent on the kinetic energy of the impactor, with impact angle and viewing geometry significantly influencing the observed rise time and peak intensity of the flash \citep{ernstEffectsViewOrientation2008}. 
High cadence light curves offer the potential to constrain these geometric parameters by resolving these features which are otherwise smoothed out in lower time resolution data. Furthermore, resolving these timescales allows for the potential separation of the initial high temperature vapor plume from the subsequent cooling ejecta cloud, providing new insights into the thermodynamics of the impact process. In this work, we present the results of a targeted survey utilising high-speed imaging systems operating at 200 and 250 FPS (5 and 4 ms between frames, respectively), which successfully confirmed four LIFs. We analyse the temporal evolution of these flashes to quantify the effects of temporal integration and investigate the physical decoupling between the initial vapour plume and the subsequent cooling ejecta.

\section{Instrumentation}
The observations were conducted using the Zadko Telescope \citep{cowardZadkoTelescopeSouthern2010, cowardZadkoTelescopeExploring2017} located approximately 70 km north of Perth, Western Australia. The telescope is an f/4 Cassegrain reflector with a primary mirror clear aperture of 1007.0 mm and a system focal length of 4038.6 mm. To achieve a wider field of view suitable for lunar monitoring, a Bintel 1.25" 0.5x focal reducer was placed in the optical path resulting in an effective focal ratio of f/2 and an effective focal length of 2019.3 mm.
\begin{figure}
    \centering
    \includegraphics[width=0.5\linewidth]{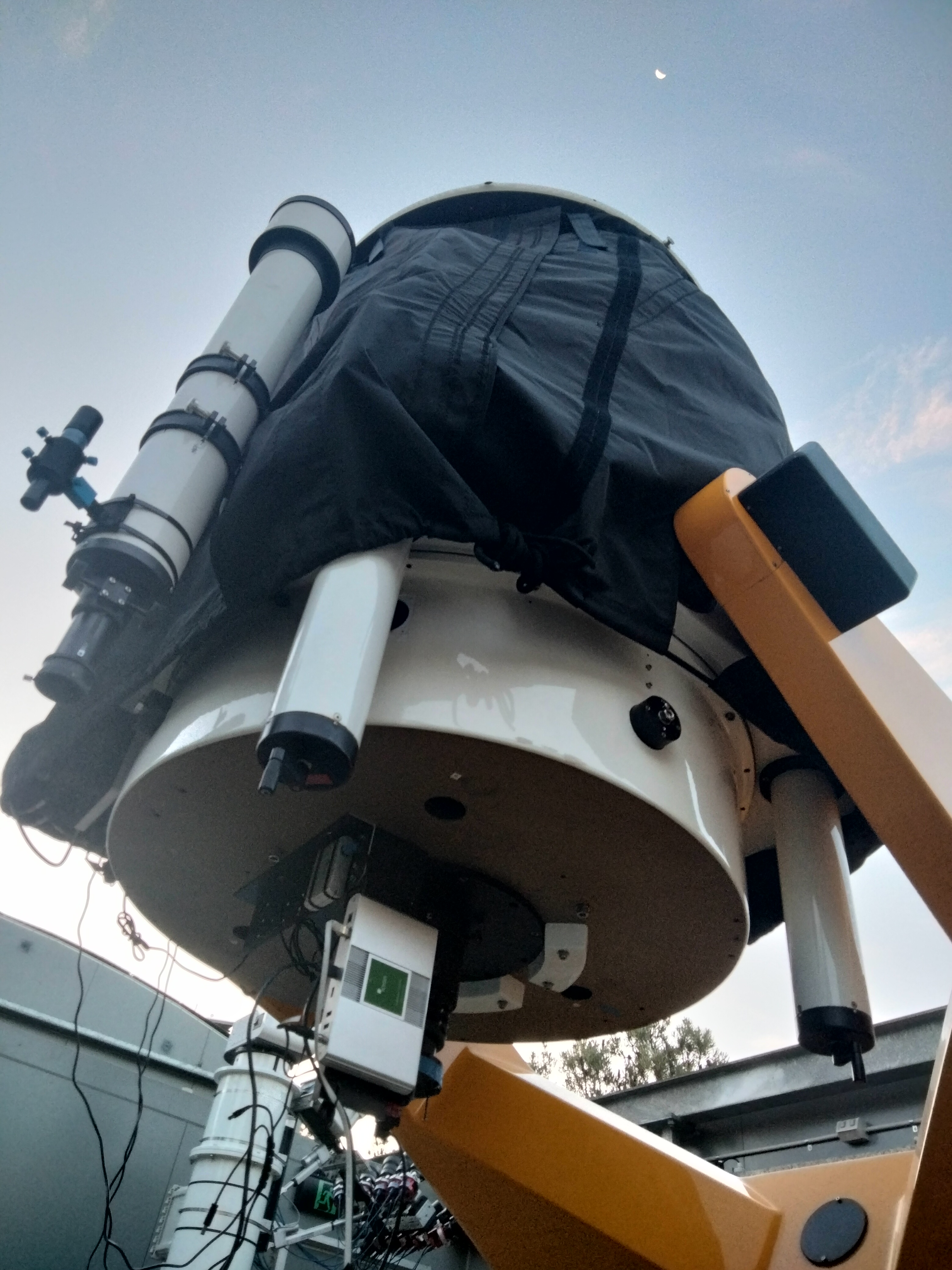}
    \caption{The Zadko telescope.}
    \label{fig:zadkoTelescope}
\end{figure}
We observe in the near-infrared bandpass to maximize the Signal-to-Noise Ratio (SNR), as LIFs are intrinsically brightest here, and the scattered Earthshine background is significantly reduced \citep{liakosNELIOTAMethodsStatistics2020,bouleyPowerDurationImpact2012}. To exploit this twofold benefit, we used a MidOpt LP715 Longpass filter with a 50\% transmission cut on wavelength at 715 nm. The detector used is an Allied Vision Alvium 1800 U-240M, which features a Sony IMX392 CMOS sensor. This camera was selected as a cost effective solution that includes features such as sensor binning capability, high frame rates, low temporal dark noise, high absolute sensitivity threshold, and a relatively high quantum efficiency in the NIR wavelengths. The USB interface also made observations simple, as they were able to run directly off consumer grade computers.
\begin{figure}
    \centering
    \includegraphics[width=0.6\linewidth]{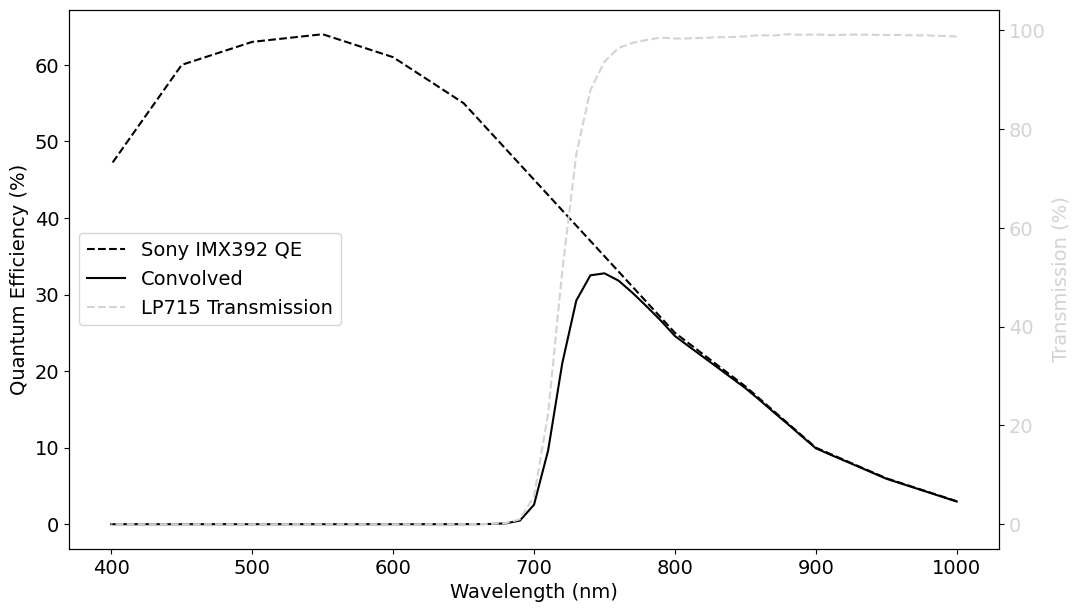}
    \caption{Spectral response of the instrumentation showing the Sony IMX392 quantum efficiency (dashed black), the LP715 filter transmission (dashed grey), and the resulting convolved system response (solid black). The convolved curve represents the effective sensitivity of the system and highlights the near-infrared bandpass used for the observations.}
    \label{fig:systemResponse}
\end{figure}

One of the flashes was also recorded by chance during an independent observation session from a second station. The equipment utilised for this independent observation consisted of a Watec 910HX RC camera mounted to a Schmidt Cassegrain telescope with a 280 mm aperture and 2800 mm focal length. A 0.33x focal reducer was employed to reduce the effective focal length to approximately 924 mm.

\section{Observations and Data Reduction}

Prior to the main campaigns a preliminary test program was conducted throughout 2023 and 2024 to validate the high-speed instrumentation at the Zadko Observatory. LIF observations were then conducted on July 29, 30 and 31, 2025 during windows selected to coincide with the favourable geometry and activity of the Southern Delta Aquariids (SDA) and Capricornids (CAP) meteor streams. A subsequent campaign was performed on December 12th and 13th, 2025, which was chosen due to the favourable geometry of the Geminids (GEM) meteor stream. December 14th should have also been a good night for Geminids LIFs, however a lightning storm on that day caused a power outage at the observatory. Across these periods the Moon was observed for a total of 7.3 hours using the observational parameters detailed in Table \ref{tab:observation_parameters}.

\begin{table}
\caption{Observation Parameters for Zadko Observatory.}
\centering
\begin{tabular}{ccccc}
\toprule
\textbf{Date} & \textbf{Start (UTC)} & \textbf{End (UTC)} & \textbf{Duration (h)} & \textbf{Moon Illumination (\%)} \\
\midrule
2025-07-29 & 10:10 & 11:30 & 1.3 & 22 \\
2025-07-30 & 10:10 & 12:30 & 2.0 & 31 \\
2025-07-31 & 10:10 & 13:20 & 2.0 & 40 \\
2025-12-12 & 19:20 & 20:40 & 1.0 & 40 \\
2025-12-13 & 19:20 & 20:40 & 1.0 & 31 \\
\bottomrule
\end{tabular}
\label{tab:observation_parameters}
\end{table}

\begin{figure}
\centering
\begin{tikzpicture}[node distance=2cm]

% Nodes
\node (start) [startstop] {Event Detected};
\node (dec1) [decision, below of=start, yshift=-0.5cm] {Spatial Movement?};
\node (process1) [process, below of=dec1, yshift=-0.5cm] {Perform Aperture Photometry};
\node (dec2) [decision, below of=process1, yshift=-0.5cm] {$\ge$ 2 Frames with SNR $>$ 5?};
\node (dec3) [decision, below of=dec2, yshift=-0.8cm] {Fast Rise Slow Decay?};
\node (dec4) [decision, below of=dec3, yshift=-0.8cm] {SatChecker: Satellite within $1^\circ$?};
\node (stop) [startstop, below of=dec4, yshift=-0.5cm] {Confirmed LIF};

% Rejection Nodes
\node (rej1) [process, right of=dec1, xshift=3.5cm, fill=gray!30] {Reject (Satellite)};
\node (rej2) [process, right of=dec2, xshift=3.5cm, fill=gray!30] {Reject (Cosmic Ray/Other)};
\node (rej3) [process, right of=dec3, xshift=3.5cm, fill=gray!30] {Reject (Noise / Other)};
\node (rej4) [process, right of=dec4, xshift=3.5cm, fill=gray!30] {Possible MEO/GEO Satellite};

% Arrows
\draw [arrow] (start) -- (dec1);
\draw [arrow] (dec1) -- node[anchor=south] {Yes} (rej1);
\draw [arrow] (dec1) -- node[anchor=east] {No} (process1);

\draw [arrow] (process1) -- (dec2);
\draw [arrow] (dec2) -- node[anchor=south] {No} (rej2);
\draw [arrow] (dec2) -- node[anchor=east] {Yes} (dec3);

\draw [arrow] (dec3) -- node[anchor=south] {No} (rej3);
\draw [arrow] (dec3) -- node[anchor=east] {Yes} (dec4);

\draw [arrow] (dec4) -- node[anchor=south] {Yes} (rej4);
\draw [arrow] (dec4) -- node[anchor=east] {No} (stop);

\end{tikzpicture}
\caption{Flowchart describing the LIF validation process.}
\label{fig:lif_flowchart}
\end{figure}
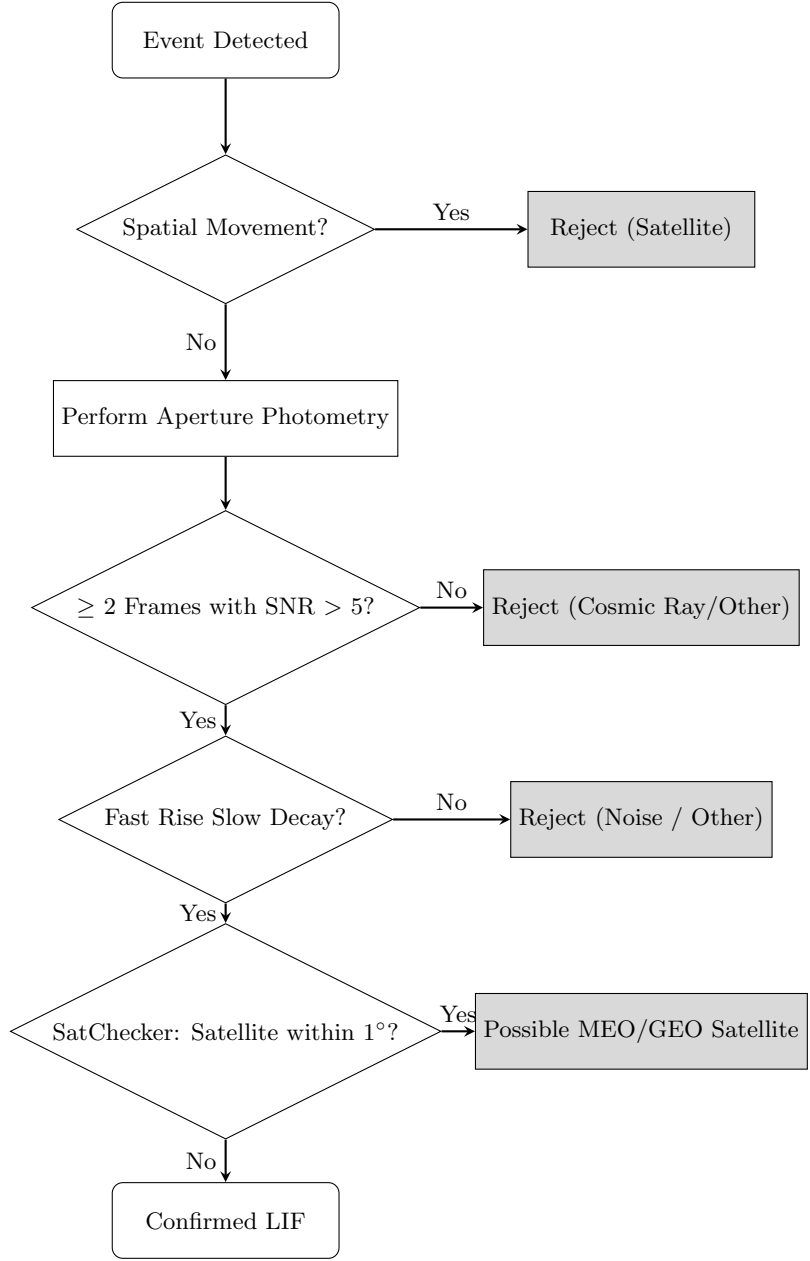

During observations of the Moon, the telescope was aimed as far away from the terminator as possible but so the limb of the Moon was still visible to cover as much surface area as possible. While lunar tracking was enabled, we had to periodically readjust the telescope every 15 minutes to ensure the limb of the Moon was still in the field of view. Network Time Protocol timing of individually received frames was used for all observations at Zadko observatory. 
For the initial observations conducted on 2025 July 29, 30, and 31, the camera was operated in the Mono12 pixel format with 2x2 sensor binning and a high gain of 47.9 dB along with a short exposure time of 4797.58 $\mu$s with 200.515 $\mu$s dedicated to image readout yielding a frame rate of 200 frames per second (FPS). This configuration produced an image size of $13.7\times8.59$ arcminutes and a pixel scale of 0.847 arcseconds per pixel after binning. At the beginning of these observations a star field was imaged for the purposes of calibration with an exposure period of 50 ms and a gain of 47.9 dB. For the entire July campaign, photometric calibration was derived from the reference star field image acquired on July 30. Photometric magnitudes were corrected for atmospheric extinction to account for the difference in airmass between the calibration stars and the location of the Moon at the time of the impacts. 
A subsequent observation campaign was conducted on 2025 December 12 and 13, with a slightly modified setup where the camera was operated with a lower gain of 40 dB and an exposure time of 3798 μs yielding a frame rate of 250 FPS. The calibration images for these nights were acquired with a gain of 35 dB and an exposure time of 2298 $\mu$s on December 13 and the same extinction correction method was applied. We recognise that utilising a single calibration epoch across multiple nights introduces systematic photometric uncertainty because atmospheric transparency and seeing conditions fluctuate between observing sessions. Furthermore, the short exposure times employed for the calibration stars render the photometric solution susceptible to errors introduced by atmospheric scintillation. Our methodology is specifically optimised to resolve the rapid morphological evolution of LIFs, providing insights into their physical dynamics. This focus ensures the data is ideally suited for high-speed temporal analysis even while absolute photometric precision remains a secondary consideration for this particular study. 
On December 13, 2025, simultaneous observations were conducted independently by a second observer located in the central Perth area between 1859 and 2030 UTC. Data was captured at 25 FPS consisting of 50 interlaced fields per second with 20 ms exposures and no optical filter. Precision timing for this system is achieved by embedding Global Navigation Satellite System (GNSS) timestamps directly into the interlaced video fields.
For observations conducted at Zadko Observatory, custom software named AlviumSyncCapture\footnote{\url{https://github.com/DaleGia/AlviumSyncCapture}} was used to read and save images to disk using multi extension Flexible Image Transport System (FITS) files. These data are later processed using separate detection software called LIFDetect\footnote{\url{https://github.com/DaleGia/LIFDetect}} . LIFDetect takes the difference between image stacks before thresholding the resulting image to a user-defined sigma value. This thresholded image is then searched for contours with an area greater than a user defined value and if a contour exists a detection is flagged. For the observed events a threshold sigma of 1.5 and a contour size of 20 pixels were employed. Images before and after the event are kept in a buffer and saved as multi extension FITS and video files for verification.
To determine whether a detection is a LIF, several criteria are evaluated as illustrated in Figure \ref{fig:lif_flowchart}. The event must not have moved over its duration as movement indicates a likely low earth orbit satellite. Following confirmation that the candidate is stationary, a light curve is produced using an aperture photometry procedure with a radius equal to the full width at half maximum of the system's point spread function.
To validate the detection, we require the signal to appear in at least two frames with SNR $>$ 5, which effectively eliminates cosmic rays that typically appear on a single frame.
While we acknowledge that this requirement may exclude genuine flashes near the sensitivity limit of the system, it ensures a robust validation standard. 
It also allows us to evaluate the detection based on its light curve shape, which must exhibit a fast rise and slow decay. As a final verification, we ensure there are no medium or geostationary orbit satellites within 1 degree of the Moon using SatChecker\footnote{\url{https://github.com/iausathub/satchecker}} which accesses the CelesTrak\footnote{\url{https://celestrak.org/}} and Space-Track\footnote{\url{https://space-track.org}} TLE datasets.

\section{Results}
\begin{figure}
    \centering
\includegraphics[width=0.8\linewidth]{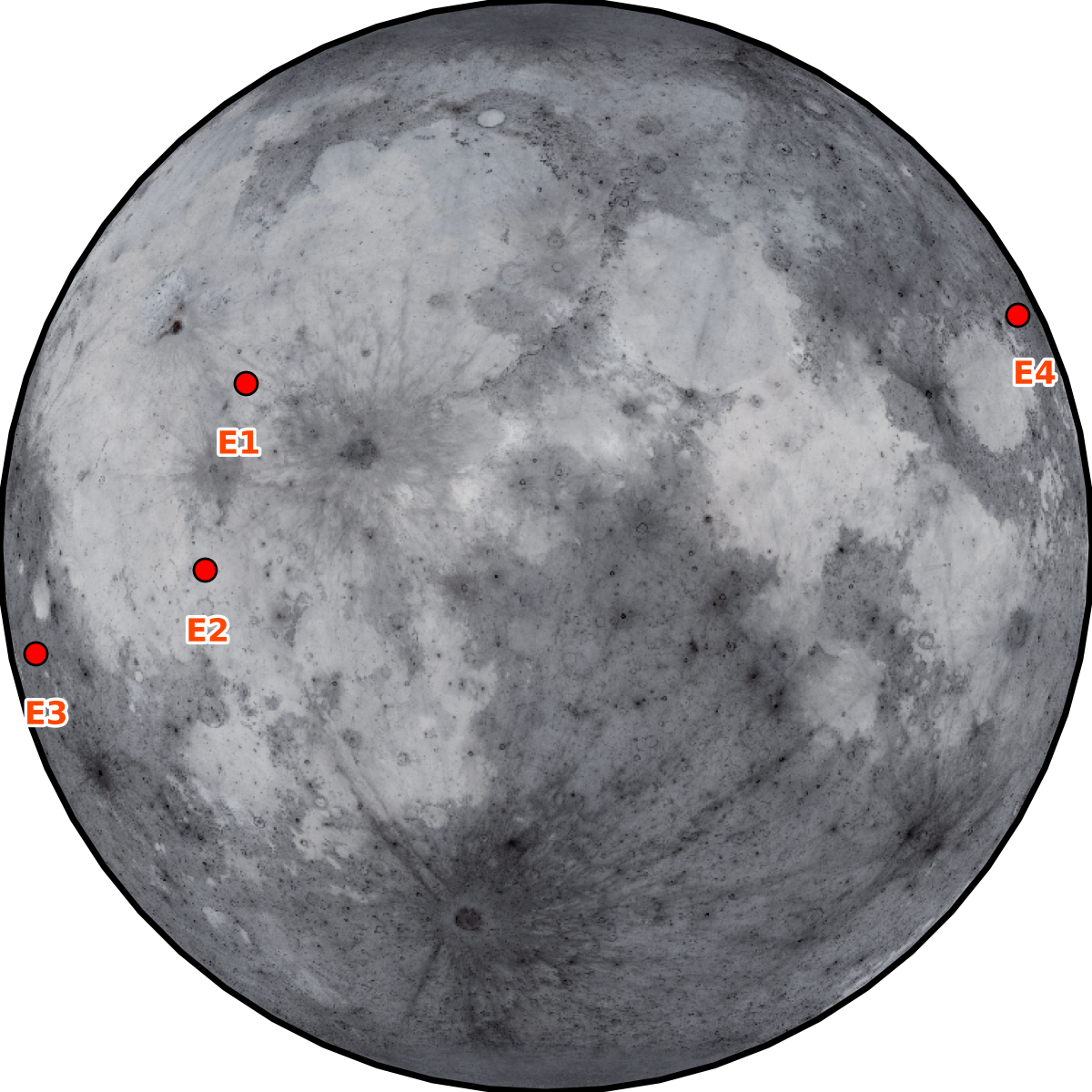}
    \caption{Impact locations of the four detected lunar flashes across both observation campaigns. The base map is generated using the LROC WAC global mosaic (Credit: NASA/GSFC/ASU).}
    \label{fig:liflocations}
\end{figure}

Across the 7.3 hours of monitored lunar observations, four bright LIFs were detected. The properties of these events are detailed in Table \ref{tab:lif_detections}, and their impact locations are visualised in Figure \ref{fig:liflocations}.

\begin{table}[ht]
\centering
\caption{Summary of observed LIFs. Integrated magnitudes are normalised to a 1.0\,s reference duration to represent total radiative output. * denotes the presence of saturated pixels resulting in underestimated magnitude values.}
\label{tab:event_summary}
\begin{tabular}{@{}llcccccccc@{}}
\toprule
\textbf{Event ID} & \textbf{Date and UT} & \textbf{$T_{dur}$ (ms)} & \textbf{FPS} & \textbf{$N_{f}$} & \textbf{Peak $I$ Mag} & \textbf{Lat ($^{\circ}$)} & \textbf{Lon ($^{\circ}$)} \\
\midrule
E1 & 2025-07-29 10:17:58.070 & 20.0 & 200 & 4 & 6.57 $\pm$ 0.12 & 17.36 & $-35.17$ \\
E2 & 2025-07-30 10:56:48.105 & 25.0 & 200 & 5 & 6.00 $\pm$ 0.11* & $-2.44$ & $-38.68$ \\
E3 & 2025-07-30 11:42:32.023 & 85.0 & 200 & 14 & 4.53 $\pm$ 0.10* & $-11.33$ & $-72.51$ \\
E4 & 2025-12-13 20:11:31.003 & 20.0 & 250 & 4 & 6.87 $\pm$ 0.07 & 25.01 & 72.23 \\
E4   & 2025-12-13 20:11:31.039 & 40.0 & 50   & 2 & 8.62 $\pm$ 0.14 &       &       \\
\bottomrule
\end{tabular}
\label{tab:lif_detections}
\end{table}

\begin{figure}[H]
    \centering
    \includegraphics[width=0.5\linewidth]{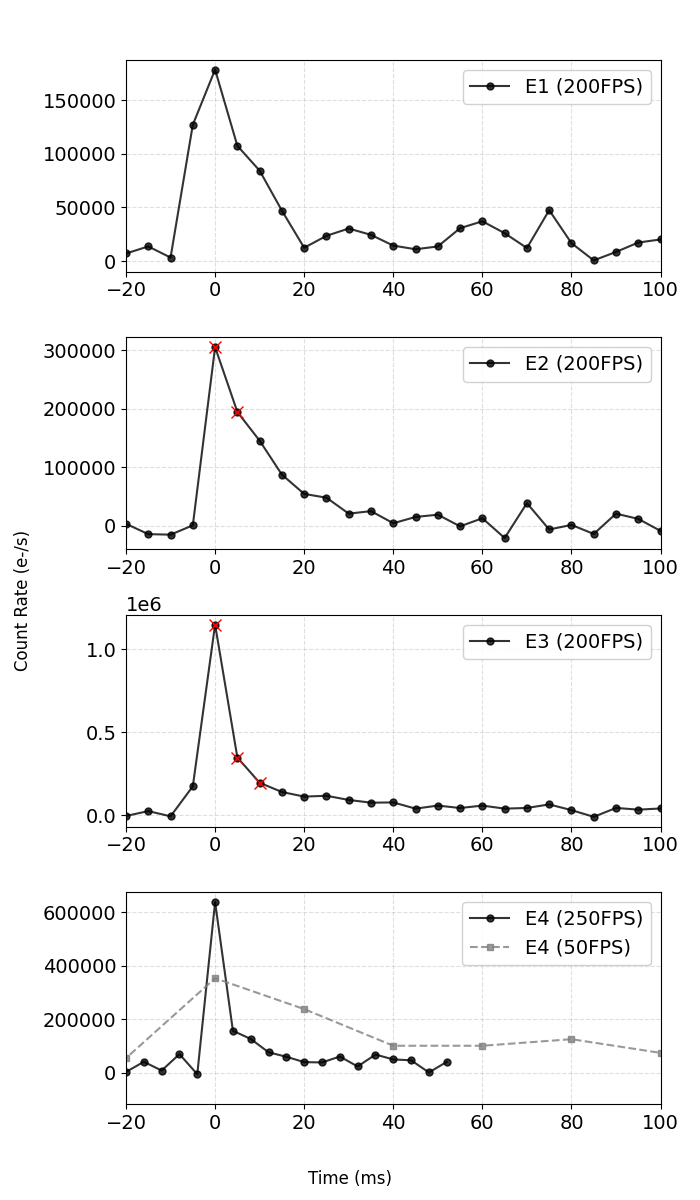}
    \caption{Light curves of each LIF recorded during the observation campaign, generated using aperture photometry. Measurements where the photometric aperture contains saturated pixels are marked with a red cross.}
    \label{fig:events_counts}
\end{figure}

\begin{figure}[H]
    \centering
    \includegraphics[width=0.5\linewidth]{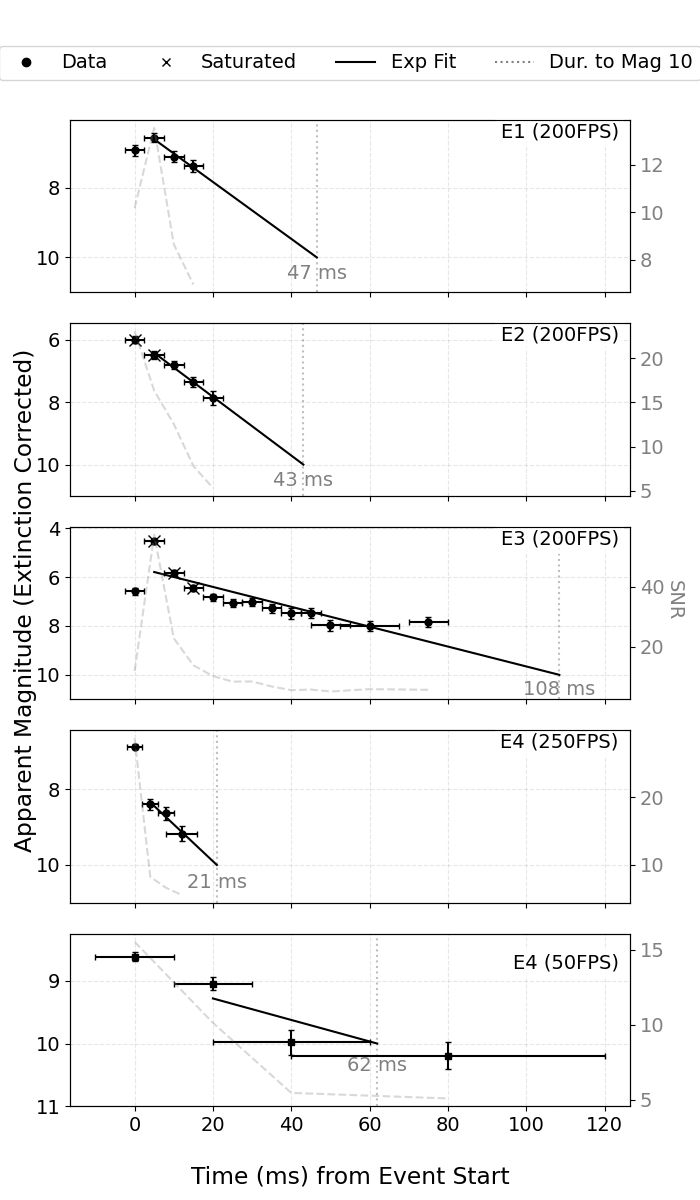}
    \caption{Extinction-corrected apparent magnitude of each LIF recorded during the observation campaign. Vertical error bars represent the combined systematic and random uncertainties in magnitude, while horizontal error bars indicate the temporal integration width determined by the adaptive binning algorithm.}
    \label{fig:events_magnitudes}
\end{figure}

\begin{figure}
\gridline{\fig{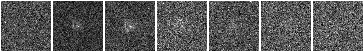}{0.8\textwidth}{(E1)}}
\gridline{\fig{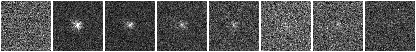}{0.8\textwidth}{(E2)}}
\gridline{\fig{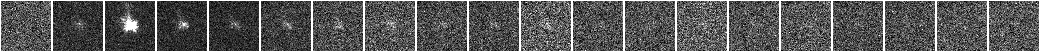}{0.8\textwidth}{(E3)}}
\gridline{\fig{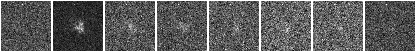}{0.8\textwidth}{(E4 (Zadko Observatory 250FPS)}}
\gridline{\fig{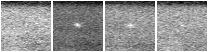}{0.8\textwidth}{(E4 50FPS)}}
\caption{Sequential image strips display the temporal evolution of the four confirmed LIFs. Panels E1 through E3 detail the high-speed captures of the first three events. The final two panels contrast the simultaneous observation of event E4 by comparing the 250 FPS data from Zadko Observatory with the 50 FPS system.}
    \label{fig:eventimages}
\end{figure}

Figure \ref{fig:events_counts} presents the light curves for each detected event while Figure \ref{fig:events_magnitudes} illustrates the extinction corrected apparent magnitudes for data points with an SNR greater than 5. Figure \ref{fig:eventimages} shows individual frames from E1, E2, E3, and E4 respectively. To increase the yield of data points meeting this SNR threshold we applied an adaptive stacking algorithm with a minimum target SNR of 5 and a maximum bin size of 50 resulting in a sensitivity gain of approximately 0.5 magnitudes for event E3 and 0.75 magnitudes for event E4. As events E2 and E3 contain saturated pixels, their reported magnitudes are underestimated. A limiting sensitivity ranging from magnitude 7.25 to 9 was recorded across the detections from Zadko Observatory where the variance was driven by the Moon's illumination phase, terminator proximity, and camera noise. The overall detection limit was restricted by noise from the uncooled camera at high gain combined with poor atmospheric conditions and optical limitations of the low cost focal reducer. Each event exhibited a characteristic fast rise followed by a slower decay with no observable spatial movement (Figure \ref{fig:events_counts}).

While spatial stability is a strong indicator of a LIF, a comprehensive check for geostationary satellites was conducted to rule out false positives. We utilised SatChecker to identify any satellites within a $1^{\circ}$ field of view of the Moon's location during a window of $\pm10$ seconds around each detection. Table \ref{tab:satellite_proximity} lists the results of this proximity analysis. Event E1 and E3 had satellites present within the search parameters. However, with a predicted minimum separation of $0.93^{\circ}$ and $0.84^{\circ}$ from the Moon, they can be disregarded as the source of the events. Finally, event E4 was simultaneously confirmed by an independent 50 FPS system operated by DH. While the observations were not coordinated, they provided validation of the detection as presented in Figure \ref{fig:eventimages}. The unfiltered system recorded a peak magnitude of 8.62, 1.75 magnitudes fainter than the peak of 6.87 measured by the 250 FPS I-band system. This discrepancy is consistent with magnitude offsets reported in previous LIF observations \citep{madiedoFirstObservationsDetermine2018}. The light curves also exhibited distinct decay profiles where the high temporal resolution system resolved a sharp decline of 1.5 magnitudes within the first 4 ms while the slower unfiltered system observed a more gradual decay of only 0.5 magnitudes over 20 ms, a discrepancy that cannot be attributed solely to the difference between filters. While the more rapid decay of I-band flux relative to the broadband signal could imply a rising temperature, the observation is consistent with the effects of temporal integration. In this scenario, the lower frame rate system effectively averages the transient, high-temperature vapour plume with the cooler, early-stage ejecta cloud.

To account for the differing sensitivity limits of each observation, we estimate the total flash duration by modelling the light curve decay and calculating the time required for the signal to fade to a limiting magnitude of 10. An exponential function was fitted to the photometric data to characterise this profile but obtaining a reliable fit for the complete dataset proved challenging for all events except E2, as the initial brightness values frequently deviated from the model. Consequently we applied the exponential fit exclusively to the data points following the first frame, which yielded a robust approximation of the decay phase. Although the function generally provided a reliable fit once the initial data point was excluded, the model struggled to accurately describe the brightest event E3. This particular flash exhibited a secondary brightness increase following an initial dimming; a behaviour consistent with features observed in other impact flashes \citep{madiedoLargeLunarImpact2014, avdellidouTemperaturesLunarImpact2019}.

Furthermore, a discrepancy was noted in the total duration for E3 where the \textit{I-band} fit estimated an end time of 21 ms while the wide band observation remained at magnitude 9 after 40 ms. This difference could be attributable to differences in the spectral response between the systems. It is important to note that the longer integration time of the 50 FPS system introduces significant uncertainty regarding the precise start and end times of the flash, as the event could commence or cease well within a single exposure window, implying that the observed duration discrepancy may be less pronounced in practice.

\begin{table}[h!]
\caption{Satellite proximity analysis.}
\centering
\begin{tabular}{cccccc}
\toprule
\textbf{Event} & \textbf{Time (UTC)} & \textbf{Satellite Name} & \textbf{NORAD} & \textbf{Dist ($^{\circ}$)} & \textbf{RA/DEC ($^{\circ}$)} \\
\midrule
E1  & 2025-07-29 10:17:58 & RADUGA 1-2 & 21038 & 0.93 & 182.14 / -1.51 \\
E2 & 2025-07-30 10:56:48 & \textit{None within $1^{\circ}$} & -- & -- & -- \\
E3 & 2025-07-30 11:42:32 & STARLINK-6332 & 57339 & 0.84 & 191.70 / -7.83 \\
E4 & 2025-12-13 20:11:31 & \textit{None within $1^{\circ}$} & -- & -- & -- \\
\bottomrule
\end{tabular}
\label{tab:satellite_proximity}
\end{table}

\begin{figure}[H]
    \centering
\includegraphics[width=0.8\linewidth]{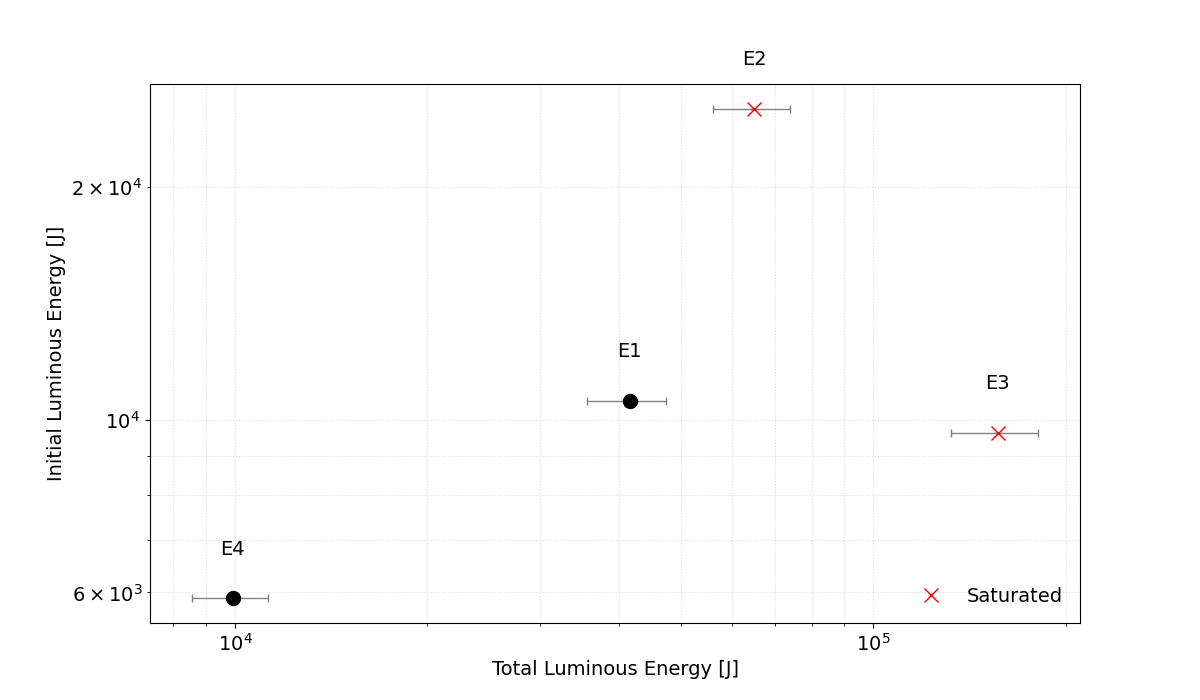}
    \caption{Initial  luminous energy and total luminous energy of each event. Both have been corrected for observation angle.}
    \label{fig:init_vs_lum}
\end{figure}

Of the four flashes detected, E1 and E3 exhibited a peak magnitude that did not occur on the first frame. This behaviour is consistent with a multiframe flash observed during the 1999 Leonids \citep{yanagisawaLightcurves1999Leonid2002} and a similar event reported by \citet{avdellidouTemperaturesLunarImpact2019}. 
Although missing light curve segments due to readout latency could explain this observation, it remains unlikely. The rapid $200~\mu s$ readout time represents a dead time of only 4–5\% and the recurrence of this feature in two of the four detected events points toward a physical origin rather than a technical artifact.
Variable atmospheric seeing offers one potential explanation, however, the rapid two magnitude brightening observed between the first and second frames of E3 renders this unlikely particularly given that saturation in the second frame implies the true flux increase was even larger.
Alternatively, we investigate if the observations resolve the physical rise time of the impact flash. As established by \citet{eichhornAnalysisHypervelocityImpact1976} this duration scales directly with projectile mass and inversely with impact velocity. However, since laboratory measurements constrain this phase to the microsecond regime for small masses, it is improbable that the rise time would extend to the millisecond scales required for detection by our system. This hypothesis is further challenged by the presence of this feature in both the bright $m_I$=4.53 E3 and the fainter $m_I$=6.57 E1, as the latter presumably involves a smaller impactor unlikely to generate such a prolonged rise phase.
A further possibility is that imaging at frame rates exceeding 200 FPS enables the temporal separation of the initial vapour plume from the subsequent incandescent ejecta.
Under this hypothesis we would expect a physical correlation to exist between the intensity of the initial phase and the total integrated energy as both originate from the same impact event. We investigated this relationship in Figure \ref{fig:init_vs_lum}, which plots the initial luminous energy against the total luminous energy. The total luminous energy was calculated following the methodology of \citet{liakosNELIOTAMethodsStatistics2020} while the initial luminous energy is defined as the energy derived solely from the first recorded data point.
To account for geometric foreshortening, we corrected these values for the observational angle bias as described by \citet{ernstEffectsViewOrientation2008} using the relation
$ E_{phys} = \frac{E_{obs}}{\sin(\theta_{obs})}$,
where $E_{obs}$ represents the observed luminous energy in Joules (J) and $\theta_{obs}$ is the observation angle. Contrary to expectations, there appears to be no discernible correlation.

\begin{figure}[H]
    \centering
\includegraphics[width=0.8\linewidth]{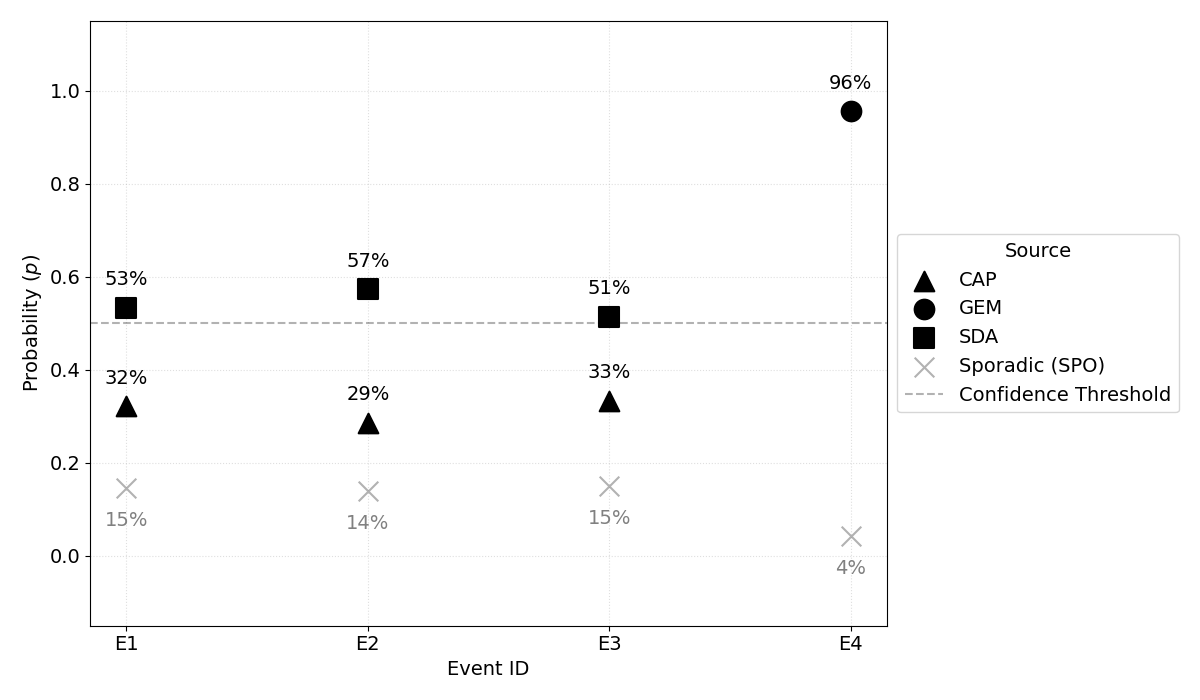}
    \caption{Estimated impactor stream source as per \citep{madiedoAnalysisMoonImpact2015}.}
\label{fig:midas_probabilities}
\end{figure}

\begin{figure}[H]
    \centering
    \includegraphics[width=0.8\linewidth]{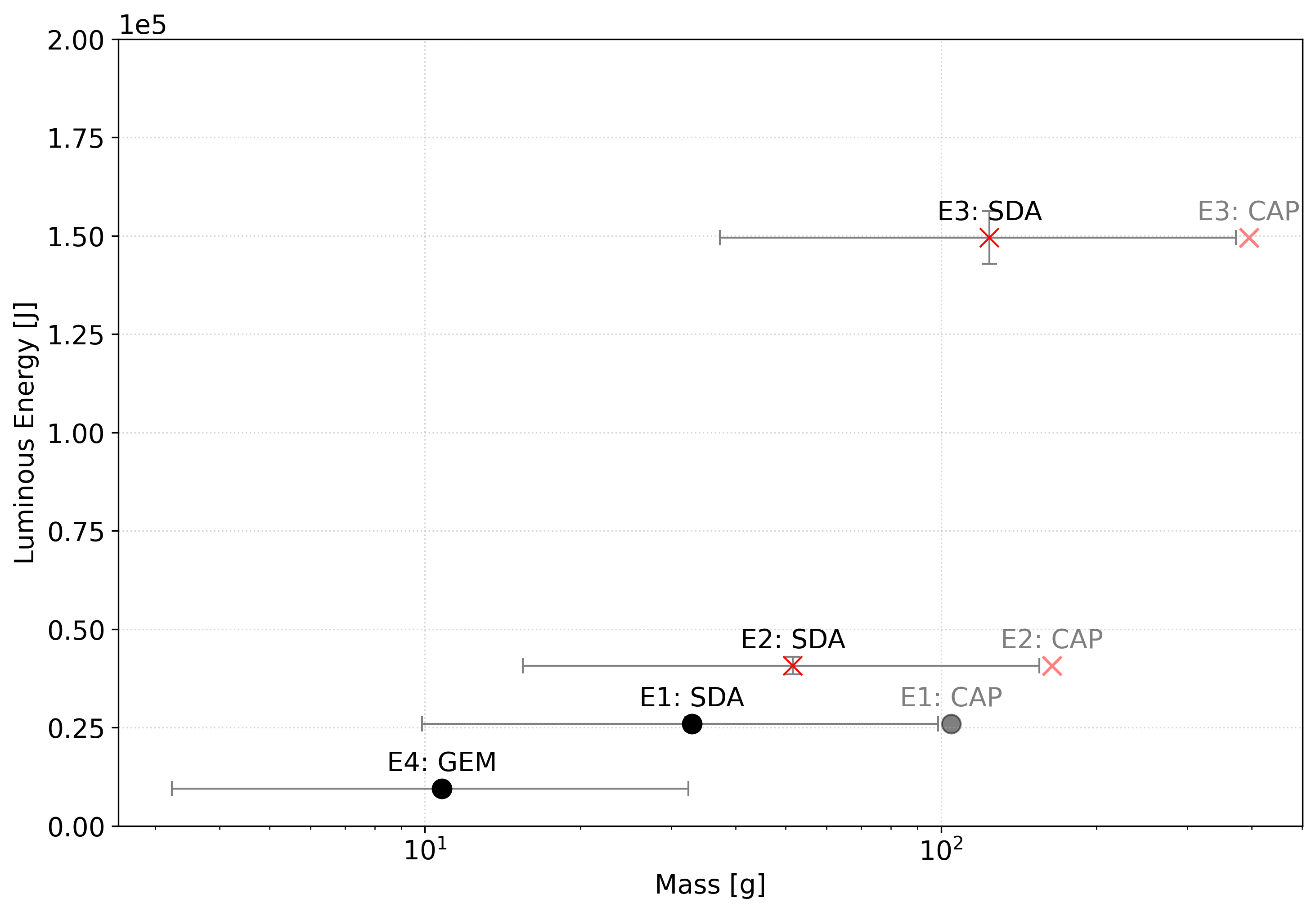}
    \caption{Luminous energy versus mass for each observed lunar impact event. Horizontal error bars represent the mass range derived from a minimum luminous efficiency ($\eta_{min} = 5 \times 10^{-4}$) and maximum efficiency ($\eta_{max} = 5 \times 10^{-3}$), and nominal luminous efficiency  ($\eta_{nom} = 1.5 \times 10^{-3}$). Vertical error bars represent the integrated photometric uncertainty. While the primary analysis assumes E1, E2, and E3 originated from a high-velocity SDA source ($v = 41$\,km/s), alternate realisations are plotted in as if they were of CAP origin ($v = 23$\,km/s). For comparison, alternate mass estimates calculated using the updated luminous efficiency of $\eta = 6.0 \times 10^{-3}$ \citep{shewardDetectionSmallFresh2025} are detailed in Table \ref{tab:lif_mass}.}
    \label{fig:lum_vs_mass}
\end{figure}

To identify the likely source of each impactor, we conducted a probabilistic analysis following the methodology of \citet{madiedoAnalysisMoonImpact2015}. As illustrated in Figure \ref{fig:midas_probabilities}, the results indicate a Southern Delta Aquariids origin for events E1 through E3, while event E4 is attributed to the Geminids stream. The specific physical and statistical parameters utilised for these calculations are detailed in Table \ref{tab:stream_parameters}.
We subsequently calculated the mass as per \citet{suggsFluxKilogramSizedMeteoroids2014} and compared it to the total luminous energy for all potential source permutations of the first three events while maintaining the Geminids assumption for E4 as illustrated in Figure \ref{fig:lum_vs_mass}.   Table \ref{tab:lif_mass} details the physical parameters for each event assuming Southern Delta Aquariids origins for the July detections and a Geminids source for the December observation.

\begin{figure}[H]
    \centering
\includegraphics[width=0.8\linewidth]{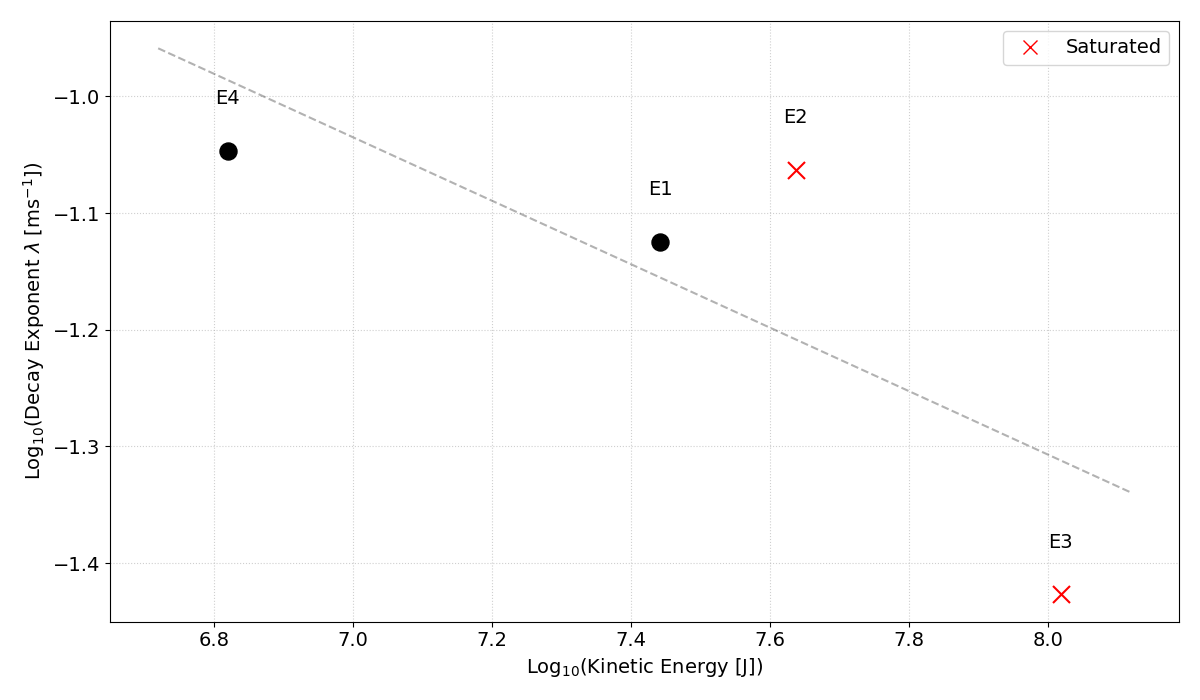}
    \caption{Decay rate vs kinetic energy for each event.}
\label{fig:decay_vs_ke}
\end{figure}

Figure \ref{fig:decay_vs_ke} presents the decay rate of each impact plotted against kinetic energy with the initial data point excluded from the analysis. As events E1, E2, and E4 exhibited comparable decay rates, it is difficult to discern a clear correlation between decay rate and kinetic energy. Establishing this relationship, which is likely driven by the changing temperature of the impacts, will require a more extensive dataset of high-cadence observations.

\begin{figure}[H]
    \centering
\includegraphics[width=0.6\linewidth]{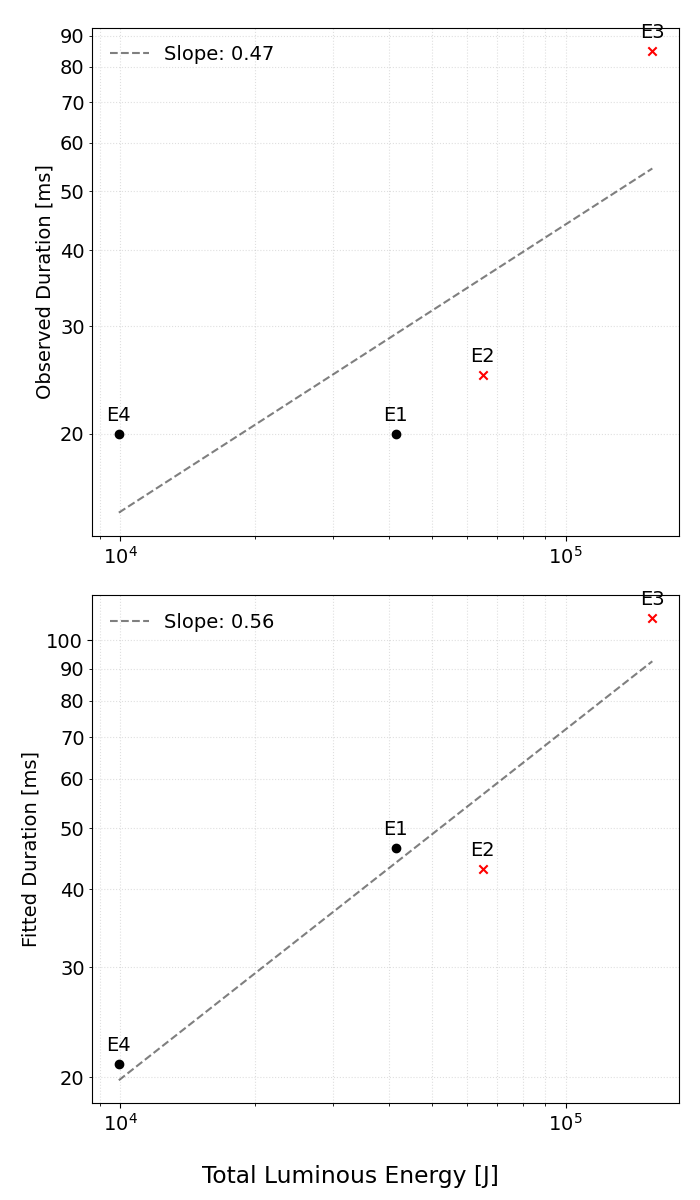}
    \caption{Comparison between the total luminous energy and the flash duration for the observed lunar impact events.}
\label{fig:duration_analysis_vertical}
\end{figure}

We observed a correlation between flash duration and luminous energy as seen in Figure \ref{fig:duration_analysis_vertical}, in agreement with the findings of \citet{bouleyPowerDurationImpact2012}. Extrapolating system sensitivity to magnitude 10 reveals that flash duration exceeded 20 milliseconds for each detected event, although all events were relatively bright.
Accurate duration measurements are hampered by the poor noise performance of the system. Without fitting the curves to a magnitude 10 threshold, three of the four events display similar observed durations. However, this remains an unfair comparison due to the differing system sensitivities and frame rates between the detection setups. When utilising the fitted duration to normalise the sensitivity limit, a clearer correlation between luminous energy and flash duration emerges. Further data involving fainter flashes with shorter durations is required to confirm this relationship.

\section{Discussion}

A key finding is that the initial luminous energy, represented by the first data point, varies much less than the total luminous energy. While the initial luminous energy across all four flashes spans only a factor of four, the total energy varies by a factor of nearly twenty. For example, E4 and E1 had very similar initial luminous energies, yet E1 produced four times as much total luminous energy.
Similarly, E4 and E3 had comparable initial luminous energies, but the total energy of E3 was over ten times higher. We checked if the impact or viewing geometry caused this, as all luminous energy values were corrected for observation angle. E1 and E2 share similar observation angles of $\sim 39^{\circ}$ defined as the angle between the local surface normal at the impact site and the observer line of sight. E3 and E4 also group together at $\sim 73^{\circ}$ yet this shows no clear pattern which could indicate that more data is needed to isolate any geometric effects.

Additionally, while E1, E3, and E4 had impact angles between $56^{\circ}$ and $65^{\circ}$, E2 struck at a much steeper $85^{\circ}$, but this difference also fails to explain the initial luminous energy similarities. Ultimately, we find no correlation between these geometric factors and the relationship between the initial and total luminous energy, suggesting more data is needed to understand why the initial plume intensity varies so differently from the total flash energy.

This physical decoupling is further corroborated by the simultaneous detection of Event E4, which provides evidence that standard video frame rates may underestimate the initial flux of LIFs. While we recorded a 1.75~magnitude discrepancy between the two systems, which is attributable to the spectral mismatch between the \textit{I-band} response of the high-speed instrument and the unfiltered response of the verification camera, the critical evidence for temporal integration lies in the light curve morphology. The high-speed system resolved a rapid luminosity drop of 1.5 magnitudes within just 4 ms, whereas the 50~FPS system recorded a much more gradual decline of 0.5~magnitudes over 20~ms. This discrepancy could also exist due to a rapid change in temperature given the briefness of the vapour plume. We note that given the microsecond-scale rise times predicted in laboratory experiments, it is probable that significant temporal integration of the peak luminosity persists even at the 200 and 250~Hz rates employed in this study. Furthermore, the absence of a correlation between the initial luminous energy and the total luminous energy suggests that the physical mechanism driving the initial vapour expansion could be physically decoupled from the longer-duration glow driven by the cooling ejecta. To fully characterise this decoupling, a larger dataset of impacts with well-constrained velocities and impact angles is required. Future campaigns should prioritise frame rates exceeding 500 FPS, which would allow for a clearer separation of the vapour and ejecta phases. Additionally, following the recommendations of \citep{bonanosNELIOTAFirstTemperature2018}, such campaigns should employ simultaneous imaging in both $I-$ and $R$-bands to further constrain the temperature evolution and physical mechanisms governing these distinct phases.

Beyond revealing magnitude discrepancies, the high temporal resolution of this study highlights the limitations of applying a simple exponential decay model to the full duration of LIFs. We find that a single exponential function fails to capture the rapid evolution of the onset. However, a reliable fit is achieved for the cooling phase once the initial data point representing the distinct vapour plume is excluded. A notable exception to this trend is observed in the most energetic event, E3. While an exponential model broadly describes the decay, the light curve displays significant deviations, characterised by phases of secondary brightening and dimming during the cool-down period. These fluctuations may indicate the presence of complex thermodynamic processes, such as a self-quenching mechanism within the expanding ejecta cloud where opacity variations modulate the observed luminosity. Although current data suggests these non-linear behaviours are characteristic of larger impact events, a more extensive catalogue of high-magnitude flashes is required to definitively quantify this effect.

We further investigated the thermodynamic evolution of the impacts by analysing the decay rate of the light curves as a function of kinetic energy. As detailed in Figure \ref{fig:decay_vs_ke}, we excluded the initial data point to isolate the cooling phase of the ejecta. Events E1, E2, and E4 exhibited remarkably similar decay rates despite their differing kinetic energies, preventing the identification of any statistically significant correlation between these parameters. This uniformity suggests that the cooling timescale may be governed more by thermal properties of the impact flashes, rather than by the total energy of the impactor itself. Alternatively, the lack of correlation may indicate that the cooling profile is driven by the changing temperature of the ejecta cloud, which may follow a consistent evolutionary track for impacts within this energy regime. A significantly larger dataset is required to determine if a clear dependence on kinetic energy emerges for a wider population of impactors.

Acquiring such a dataset remains a challenge as the absence of fainter detections in this survey is likely attributable to a combination of insufficient sensitivity to resolve low-SNR events and the restricted field of view, which limits the statistical probability of capturing these more frequent impacts. The high gain settings employed during the initial July 2025 observations resulted in pixel saturation and the underestimation of peak brightness for the detected events. While future campaigns would benefit from low-noise cooled cameras to address sensitivity, expanding the monitoring area while maintaining high frame rates presents a significant technical challenge. Achieving a large field of view without sacrificing spatial resolution requires sensors with high pixel counts, resulting in substantial data rates that complicate real-time processing and storage. Furthermore, atmospheric scintillation likely degrades the SNR and distorts light curve morphology. These random errors are potentially more significant in high frame rate data compared to standard systems, although a detailed quantification of these effects was beyond the scope of this study.

\section{Conclusion}

High-speed observations at 200 and 250~FPS reveal complex light curve morphologies that are poorly described by exponential decay models typically used in 25--60~FPS surveys. Simultaneous observations confirmed that 25--60~FPS systems underestimate the initial magnitude of these events, most likely due to the temporal integration of the short-lived vapour plume with the subsequent cooling phase. Furthermore, the lack of statistical correlation between the initial luminous energy and the total integrated energy of LIFs could indicate that the vapour expansion is physically decoupled from the ejecta cloud. This hypothesis is supported by our observation that the luminous energy of the initial vapor plume exhibits significantly less variance between events than the total luminous energy. To fully characterise this behaviour a larger dataset of impacts with well-constrained velocities and impact angles is required. Future monitoring programs should prioritise cooled low-noise cameras operating at frame rates exceeding 500~FPS, with simultaneous multi-band photometry to increase total system sensitivity to detect fainter flashes and fully constrain the thermodynamics of the impact process.

\begin{acknowledgments}
We thank Elisa Maria Alessi, Detlef Koschny, and Tony Cook for coordinating the Geminids 2025 Lunar Impact Flash observation campaign funded by Italian Space Agency, through the LUMIO agreement n. 2024-6-HH.0\footnote{\url{https://lif.mi.imati.cnr.it}}, enabling us to learn about the concurrent observations made by DH. The authors used Gemini to improve the readability and language of this manuscript. The authors reviewed and revised the output and take full responsibility for the content.
\end{acknowledgments}

\subsubsection*{Data Availability Statement}
Supplementary materials have been uploaded as a Zenodo record at \href{https://doi.org/10.5281/zenodo.18917249}{https://doi.org/10.5281/zenodo.18917249}.

\appendix

\begin{table}[ht]
\centering
\caption{Physical and statistical parameters for the relevant meteoroid streams and the sporadic background as used in this work.}
\label{tab:stream_parameters}
\begin{tabular}{@{}lccccc@{}}
\toprule
\textbf{Stream} & \textbf{Code} & \textbf{$V_p$ (km/s)} & \textbf{ZHR/HR ($h^{-1}$)} & \textbf{$r$} & \textbf{$m_0$ (kg)} \\
\midrule
(Sporadics) & SPO & 17.0 & 10.0 & 3.0 & $5.00 \times 10^{-6}$ \\
$\alpha$-Capricornids & CAP & 23.0 & 6.0 & 2.5 & $2.76 \times 10^{-6}$ \\
Geminids & GEM & 36.0 & 88.0 & 2.4 & $4.56 \times 10^{-7}$ \\
Southern $\delta$-Aquariids & SDA & 41.0 & 25.0 & 2.5 & $2.33 \times 10^{-7}$ \\
\bottomrule
\end{tabular}
\end{table}

\begin{table}[ht]
\centering
\caption{\centering Physical properties of observed LIFs derived from the SDA (E1--E3) and Geminid (E4) stream assignments. Asterisks (*) denote saturation effects likely leading to underestimated energy and mass values.}
\label{tab:lif_mass}
\begin{tabular}{@{}llcccccc@{}}
\toprule
\textbf{Event ID} & \textbf{Source} & \textbf{$V_{imp}$ (km/s)} & \textbf{$\theta_{imp}$ ($^{\circ}$)} & \textbf{$\theta_{obs}$ ($^{\circ}$)} & \textbf{$E_{k}$ ($10^6$ J)} & \textbf{Mass (g)} & \textbf{Mass \citep{shewardDetectionSmallFresh2025} (g)} \\
\midrule
E1 & SDA & 41.0 & 63.9 & 38.7 & $27.66^{+55.33}_{-19.36}$ & $32.91^{+65.82}_{-23.04}$ &         $8.23^{+2.06}_{-1.37}$ \\
E2 & SDA & 41.0 & 85.4 & 38.8 & $43.38^{+86.77}_{-30.37}$* & $51.62^{+103.24}_{-36.13}$* &      $12.90^{+3.23}_{-2.15}$* \\
E3 & SDA & 41.0 & 56.2 & 72.9 & $104.36^{+208.72}_{-73.05}$* & $124.16^{+248.32}_{-86.91}$* &   $31.03^{+7.76}_{-5.17}$* \\
E4 & GEM & 36.0 & 65.1 & 73.9 & $6.61^{+13.22}_{-4.63}$ & $10.79^{+21.58}_{-7.55}$ &            $2.70^{+0.67}_{-0.45}$ \\
\bottomrule
\end{tabular}
\end{table}

% \section{Appendix information}

\bibliography{references}{}

@inproceedings{ernstEffectsViewOrientation2008,
	title = {Effects of {View} {Orientation} on {Impact} {Flash} {Observations}: {Implications} for {Lunar} {Impacts}},
	shorttitle = {Effects of {View} {Orientation} on {Impact} {Flash} {Observations}},
	url = {https://ui.adsabs.harvard.edu/abs/2008LPI....39.2291E},
	urldate = {2026-01-09},
	booktitle = {39th {Annual} {Lunar} and {Planetary} {Science} {Conference}},
	author = {Ernst, C. M. and Schultz, P. H.},
	month = mar,
	year = {2008},
	note = {ADS Bibcode: 2008LPI....39.2291E},
	pages = {2291},
}

@inproceedings{ernstEffectInitialConditions2003,
	title = {Effect of {Initial} {Conditions} on {Impact} {Flash} {Decay}},
	url = {https://ui.adsabs.harvard.edu/abs/2003LPI....34.2020E},
	urldate = {2025-10-13},
	booktitle = {Lunar and {Planetary} {Science} {Conference}},
	author = {Ernst, C. M. and Schultz, P. H.},
	month = mar,
	year = {2003},
	note = {ADS Bibcode: 2003LPI....34.2020E},
	pages = {2020},
}

@inproceedings{ernstHighSpeedImagingImpact2010,
	title = {High-{Speed} {Imaging} of the {Impact} {Flash}: {Observations} of {Source} {Location} and {Transient} {Crater} {Growth}},
	shorttitle = {High-{Speed} {Imaging} of the {Impact} {Flash}},
	url = {https://ui.adsabs.harvard.edu/abs/2010LPI....41.1381E},
	urldate = {2026-01-09},
	booktitle = {41st {Annual} {Lunar} and {Planetary} {Science} {Conference}},
	author = {Ernst, C. M. and Barnouin, O. S. and Schultz, P. H.},
	month = mar,
	year = {2010},
	note = {ADS Bibcode: 2010LPI....41.1381E},
	pages = {1381},
}

@inproceedings{artemievaLunarLeonidMeteors2000,
	title = {Lunar {Leonid} {Meteors} - {Numerical} {Simulations}},
	url = {https://ui.adsabs.harvard.edu/abs/2000LPI....31.1402A},
	urldate = {2025-10-13},
	booktitle = {31st {Annual} {Lunar} and {Planetary} {Science} {Conference}},
	author = {Artemieva, N. A. and Shuvalov, V. V. and Trubetskaya, I. A.},
	month = mar,
	year = {2000},
	note = {ADS Bibcode: 2000LPI....31.1402A},
	pages = {1402},
}

@article{shewardDetectionSmallFresh2025,
	title = {Detection of small fresh craters on the {Moon}: {Linking} fresh craters to their lunar impact flash events},
	volume = {699},
	issn = {0004-6361},
	shorttitle = {Detection of small fresh craters on the {Moon}},
	url = {https://ui.adsabs.harvard.edu/abs/2025A&A...699L...3S},
	doi = {10.1051/0004-6361/202555481},
	urldate = {2026-04-28},
	journal = {Astronomy and Astrophysics},
	publisher = {EDP},
	author = {Sheward, D. and Delbo, M. and Avdellidou, C. and Cook, A. and Lognonné, P.},
	month = jul,
	year = {2025},
	note = {ADS Bibcode: 2025A\&A...699L...3S},
	pages = {L3},
}

@article{cowardZadkoTelescopeExploring2017,
	title = {The {Zadko} {Telescope}: {Exploring} the {Transient} {Universe}},
	volume = {34},
	issn = {1323-3580, 1448-6083},
	shorttitle = {The {Zadko} {Telescope}},
	url = {https://www.cambridge.org/core/journals/publications-of-the-astronomical-society-of-australia/article/zadko-telescope-exploring-the-transient-universe/9AD382F5363E8648388C6C2494C0FF1B},
	doi = {10.1017/pasa.2016.61},
	language = {en},
	urldate = {2026-02-23},
	journal = {Publications of the Astronomical Society of Australia},
	author = {Coward, D. M. and Gendre, B. and Tanga, P. and Turpin, D. and Zadko, J. and Dodson, R. and Devogéle, M. and Howell, E. J. and Kennewell, J. A. and Boër, M. and Klotz, A. and Dornic, D. and Moore, J. A. and Heary, A.},
	month = jan,
	year = {2017},
	pages = {e005},
}

@article{yanagisawaLowDispersionSpectra2025,
	title = {Low dispersion spectra of lunar impact flashes},
	issn = {0019-1035},
	url = {https://www.sciencedirect.com/science/article/pii/S0019103525000272},
	doi = {10.1016/j.icarus.2025.116480},
	urldate = {2025-03-06},
	journal = {Icarus},
	author = {Yanagisawa, Masahisa and Uchida, Yuki and Kurihara, Seiya and Abe, Shinsuke and Fuse, Ryota and Tanaka, Satoshi and Onodera, Keisuke and Kawamura, Taichi and Yamada, Ryuhei},
	month = jan,
	year = {2025},
	pages = {116480},
}

@article{liakosNELIOTANewResults2024,
	title = {{NELIOTA}: {New} results and updated statistics after 6.5 years of lunar impact flashes monitoring},
	volume = {687},
	copyright = {© The Authors 2024},
	issn = {0004-6361, 1432-0746},
	shorttitle = {{NELIOTA}},
	url = {https://www.aanda.org/articles/aa/abs/2024/07/aa49542-24/aa49542-24.html},
	doi = {10.1051/0004-6361/202449542},
	language = {en},
	urldate = {2025-03-06},
	journal = {Astronomy \& Astrophysics},
	publisher = {EDP Sciences},
	author = {Liakos, A. and Bonanos, A. Z. and Xilouris, E. M. and Koschny, D. and Bellas-Velidis, I. and Boumis, P. and Maroussis, A. and Moissl, R.},
	month = jul,
	year = {2024},
	pages = {A14},
}

@article{suggsFluxKilogramSizedMeteoroids2014,
	title = {The {Flux} of {Kilogram}-{Sized} {Meteoroids} from {Lunar} {Impact} {Monitoring}},
	volume = {238},
	issn = {00191035},
	url = {http://arxiv.org/abs/1404.6458},
	doi = {10.1016/j.icarus.2014.04.032},
	urldate = {2023-08-09},
	journal = {Icarus},
	author = {Suggs, Robert and Moser, Danielle and Cooke, William and Suggs, Ronnie},
	month = aug,
	year = {2014},
	pages = {23--36},
}

@article{madiedoAnalysisMoonImpact2015,
	title = {Analysis of {Moon} impact flashes detected during the 2012 and 2013 {Perseids}},
	volume = {577},
	issn = {0004-6361, 1432-0746},
	url = {http://arxiv.org/abs/1503.05227},
	doi = {10.1051/0004-6361/201525656},
	urldate = {2023-08-09},
	journal = {Astronomy \& Astrophysics},
	author = {Madiedo, José M. and Ortiz, José L. and Organero, Faustino and Ana-Hernández, Leonor and Fonseca, Fernando and Morales, Nicolás and Cabrera-Caño, Jesús},
	month = may,
	year = {2015},
	pages = {A118},
}

@article{tandyImpactFlashEvolution2020,
	title = {Impact flash evolution of {CO2} ice, water ice, and frozen {Martian} and lunar regolith simulant targets},
	volume = {55},
	copyright = {© 2020 The Authors. Meteoritics \& Planetary Science published by Wiley Periodicals LLC on behalf of The Meteoritical Society (MET)},
	issn = {1945-5100},
	url = {https://onlinelibrary.wiley.com/doi/abs/10.1111/maps.13581},
	doi = {10.1111/maps.13581},
	language = {en},
	number = {10},
	urldate = {2026-01-27},
	journal = {Meteoritics \& Planetary Science},
	author = {Tandy, Jon D. and Price, Mark C. and Wozniakiewicz, Penny J. and Cole, Mike J. and Alesbrook, Luke S. and Avdellidou, Chrysa},
	year = {2020},
	note = {\_eprint: https://onlinelibrary.wiley.com/doi/pdf/10.1111/maps.13581},
	pages = {2301--2319},
}

@article{yanagisawaLowdispersionSpectralVideo2022,
	title = {A low-dispersion spectral video camera for observing lunar impact flashes},
	volume = {74},
	issn = {1880-5981},
	url = {https://doi.org/10.1186/s40623-022-01575-9},
	doi = {10.1186/s40623-022-01575-9},
	language = {en},
	number = {1},
	urldate = {2026-01-22},
	journal = {Earth, Planets and Space},
	author = {Yanagisawa, Masahisa and Kakinuma, Fumihiro},
	month = apr,
	year = {2022},
	pages = {62},
}

@article{madiedoFirstObservationsDetermine2018,
	title = {The first observations to determine the temperature of a lunar impact flash and its evolution},
	volume = {480},
	issn = {0035-8711},
	url = {https://doi.org/10.1093/mnras/sty1862},
	doi = {10.1093/mnras/sty1862},
	number = {4},
	urldate = {2026-01-15},
	journal = {Monthly Notices of the Royal Astronomical Society},
	author = {Madiedo, José M and Ortiz, José L and Morales, Nicolás},
	month = nov,
	year = {2018},
	pages = {5010--5016},
}

@article{cowardZadkoTelescopeSouthern2010,
	title = {The {Zadko} {Telescope}: {A} {Southern} {Hemisphere} {Telescope} for {Optical} {Transient} {Searches}, {Multi}-{Messenger} {Astronomy} and {Education}},
	volume = {27},
	issn = {1448-6083, 1323-3580},
	shorttitle = {The {Zadko} {Telescope}},
	url = {https://www.cambridge.org/core/journals/publications-of-the-astronomical-society-of-australia/article/zadko-telescope-a-southern-hemisphere-telescope-for-optical-transient-searches-multimessenger-astronomy-and-education/A6CB1AC8A98F0621F45976A1A99D0100},
	doi = {10.1071/AS09078},
	language = {en},
	number = {3},
	urldate = {2025-10-22},
	journal = {Publications of the Astronomical Society of Australia},
	author = {Coward, D. M. and Todd, M. and Vaalsta, T. P. and Laas-Bourez, M. and Klotz, A. and Imerito, A. and Yan, L. and Luckas, P. and Fletcher, A. B. and Zadnik, M. G. and Burman, R. R. and Blair, D. G. and Zadko, J. and Boër, M. and Thierry, P. and Howell, E. J. and Gordon, S. and Ahmat, A. and Moore, J. A. and Frost, K.},
	month = jan,
	year = {2010},
	pages = {331--339},
}

@article{lognonneMoonMeteoriticSeismic2009,
	title = {Moon meteoritic seismic hum: {Steady} state prediction},
	volume = {114},
	copyright = {Copyright 2009 by the American Geophysical Union.},
	issn = {2156-2202},
	shorttitle = {Moon meteoritic seismic hum},
	url = {https://onlinelibrary.wiley.com/doi/abs/10.1029/2008JE003294},
	doi = {10.1029/2008JE003294},
	language = {en},
	number = {E12},
	urldate = {2025-10-13},
	journal = {Journal of Geophysical Research: Planets},
	author = {Lognonné, Philippe and Le Feuvre, Mathieu and Johnson, Catherine L. and Weber, Renee C.},
	year = {2009},
	note = {\_eprint: https://agupubs.onlinelibrary.wiley.com/doi/pdf/10.1029/2008JE003294},
}

@article{yamadaOptimisationSeismicNetwork2011,
	title = {Optimisation of seismic network design: {Application} to a geophysical international lunar network},
	volume = {59},
	issn = {0032-0633},
	shorttitle = {Optimisation of seismic network design},
	url = {https://www.sciencedirect.com/science/article/pii/S0032063310003673},
	doi = {10.1016/j.pss.2010.12.007},
	number = {4},
	urldate = {2025-10-13},
	journal = {Planetary and Space Science},
	author = {Yamada, Ryuhei and Garcia, Raphaël F. and Lognonné, Philippe and Feuvre, Mathieu Le and Calvet, Marie and Gagnepain-Beyneix, Jeannine},
	month = mar,
	year = {2011},
	pages = {343--354},
}

@article{kadonoObservationExpandingVapor1996,
	title = {Observation of expanding vapor cloud generated by hypervelocity impact},
	volume = {101},
	copyright = {Copyright 1997 by the American Geophysical Union.},
	issn = {2156-2202},
	url = {https://onlinelibrary.wiley.com/doi/abs/10.1029/96JE02795},
	doi = {10.1029/96JE02795},
	language = {en},
	number = {E11},
	urldate = {2025-10-13},
	journal = {Journal of Geophysical Research: Planets},
	author = {Kadono, Toshihiko and Fujiwara, Akira},
	year = {1996},
	note = {\_eprint: https://agupubs.onlinelibrary.wiley.com/doi/pdf/10.1029/96JE02795},
	pages = {26097--26109},
}

@article{eichhornAnalysisHypervelocityImpact1976,
	title = {Analysis of the hypervelocity impact process from impact flash measurements},
	volume = {24},
	issn = {0032-0633},
	url = {https://www.sciencedirect.com/science/article/pii/0032063376901148},
	doi = {10.1016/0032-0633(76)90114-8},
	number = {8},
	urldate = {2025-10-13},
	journal = {Planetary and Space Science},
	author = {Eichhorn, G.},
	month = aug,
	year = {1976},
	pages = {771--781},
}

@article{eichhornMeasurementsLightFlash1975,
	title = {Measurements of the light flash produced by high velocity particle impact},
	volume = {23},
	issn = {0032-0633},
	url = {https://www.sciencedirect.com/science/article/pii/0032063375900057},
	doi = {10.1016/0032-0633(75)90005-7},
	number = {11},
	urldate = {2025-10-13},
	journal = {Planetary and Space Science},
	author = {Eichhorn, G.},
	month = nov,
	year = {1975},
	pages = {1519--1525},
}

@article{yanagisawaLightcurves1999Leonid2002,
	title = {Lightcurves of 1999 {Leonid} {Impact} {Flashes} on the {Moon}},
	volume = {159},
	issn = {0019-1035},
	url = {https://www.sciencedirect.com/science/article/pii/S0019103502969319},
	doi = {10.1006/icar.2002.6931},
	number = {1},
	urldate = {2025-10-13},
	journal = {Icarus},
	author = {Yanagisawa, Masahisa and Kisaichi, Narumi},
	month = sep,
	year = {2002},
	pages = {31--38},
}

@article{ortizDetectionSporadicImpact2006,
	title = {Detection of sporadic impact flashes on the {Moon}: {Implications} for the luminous efficiency of hypervelocity impacts and derived terrestrial impact rates},
	volume = {184},
	issn = {0019-1035},
	shorttitle = {Detection of sporadic impact flashes on the {Moon}},
	url = {https://www.sciencedirect.com/science/article/pii/S0019103506001606},
	doi = {10.1016/j.icarus.2006.05.002},
	number = {2},
	urldate = {2025-10-13},
	journal = {Icarus},
	author = {Ortiz, J. L. and Aceituno, F. J. and Quesada, J. A. and Aceituno, J. and Fernández, M. and Santos-Sanz, P. and Trigo-Rodríguez, J. M. and Llorca, J. and Martín-Torres, F. J. and Montañés-Rodríguez, P. and Pallé, E.},
	month = oct,
	year = {2006},
	pages = {319--326},
}

@article{ortizOpticalDetectionMeteoroidal2000,
	title = {Optical detection of meteoroidal impacts on the {Moon}},
	volume = {405},
	copyright = {2000 Macmillan Magazines Ltd.},
	issn = {1476-4687},
	url = {https://www.nature.com/articles/35016015},
	doi = {10.1038/35016015},
	language = {en},
	number = {6789},
	urldate = {2025-10-13},
	journal = {Nature},
	publisher = {Nature Publishing Group},
	author = {Ortiz, J. L. and Sada, P. V. and Bellot Rubio, L. R. and Aceituno, F. J. and Aceituno, J. and Gutiérrez, P. J. and Thiele, U.},
	month = jun,
	year = {2000},
	pages = {921--923},
}

@article{liakosNELIOTAMethodsStatistics2020,
	title = {{NELIOTA}: {Methods}, statistics, and results for meteoroids impacting the {Moon}},
	volume = {633},
	copyright = {© ESO 2020},
	issn = {0004-6361, 1432-0746},
	shorttitle = {{NELIOTA}},
	url = {https://www.aanda.org/articles/aa/abs/2020/01/aa36709-19/aa36709-19.html},
	doi = {10.1051/0004-6361/201936709},
	language = {en},
	urldate = {2023-09-18},
	journal = {Astronomy \& Astrophysics},
	publisher = {EDP Sciences},
	author = {Liakos, A. and Bonanos, A. Z. and Xilouris, E. M. and Koschny, D. and Bellas-Velidis, I. and Boumis, P. and Charmandaris, V. and Dapergolas, A. and Fytsilis, A. and Maroussis, A. and Moissl, R.},
	month = jan,
	year = {2020},
	pages = {A112},
}

@article{avdellidouImpactsMoonAnalysis2021,
	title = {Impacts on the {Moon}: {Analysis} methods and size distribution of impactors},
	volume = {200},
	issn = {0032-0633},
	shorttitle = {Impacts on the {Moon}},
	url = {https://www.sciencedirect.com/science/article/pii/S0032063321000404},
	doi = {10.1016/j.pss.2021.105201},
	urldate = {2023-09-18},
	journal = {Planetary and Space Science},
	author = {Avdellidou, Chrysa and Munaibari, Edhah and Larson, Raven and Vaubaillon, Jeremie and Delbo, Marco and Hayne, Paul and Wieczorek, Mark and Sheward, Daniel and Cook, Antony},
	month = jun,
	year = {2021},
	pages = {105201},
}

@article{bonanosNELIOTAFirstTemperature2018,
	title = {{NELIOTA}: {First} temperature measurement of lunar impact flashes},
	volume = {612},
	copyright = {© ESO 2018},
	issn = {0004-6361, 1432-0746},
	shorttitle = {{NELIOTA}},
	url = {https://www.aanda.org/articles/aa/abs/2018/04/aa32109-17/aa32109-17.html},
	doi = {10.1051/0004-6361/201732109},
	language = {en},
	urldate = {2023-09-18},
	journal = {Astronomy \& Astrophysics},
	publisher = {EDP Sciences},
	author = {Bonanos, A. Z. and Avdellidou, C. and Liakos, A. and Xilouris, E. M. and Dapergolas, A. and Koschny, D. and Bellas-Velidis, I. and Boumis, P. and Charmandaris, V. and Fytsilis, A. and Maroussis, A.},
	month = apr,
	year = {2018},
	pages = {A76},
}

@misc{liakosNELIOTALunarImpact2019,
	title = {{NELIOTA} {Lunar} {Impact} {Flash} {Detection} and {Event} {Validation}},
	url = {http://arxiv.org/abs/1901.11414},
	doi = {10.48550/arXiv.1901.11414},
	urldate = {2023-08-22},
	publisher = {arXiv},
	author = {Liakos, Alexios and Bonanos, Alceste and Xilouris, Emmanouil and Bellas-Velidis, Ioannis and Boumis, Panayotis and Charmandaris, Vassilis and Dapergolas, Anastasios and Fytsilis, Anastasios and Maroussis, Athanassios and Koschny, Detlef and Moissl, Richard and Navarro, Vicente},
	month = jan,
	year = {2019},
}

@incollection{suggsNASALunarImpact2008,
	address = {New York, NY},
	title = {The {NASA} {Lunar} {Impact} {Monitoring} {Program}},
	isbn = {978-0-387-78419-9},
	url = {https://doi.org/10.1007/978-0-387-78419-9_41},
	language = {en},
	urldate = {2023-08-09},
	booktitle = {Advances in {Meteoroid} and {Meteor} {Science}},
	publisher = {Springer},
	author = {Suggs, Robert M. and Cooke, William J. and Suggs, Ronnie J. and Swift, Wesley R. and Hollon, Nicholas},
	editor = {Trigo-Rodríguez, J. M. and Rietmeijer, F. J. M. and Llorca, J. and Janches, D.},
	year = {2008},
	pages = {293--298},
}

@article{avdellidouTemperaturesLunarImpact2019,
	title = {Temperatures of lunar impact flashes: mass and size distribution of small impactors hitting the {Moon}},
	volume = {484},
	issn = {0035-8711},
	shorttitle = {Temperatures of lunar impact flashes},
	url = {https://doi.org/10.1093/mnras/stz355},
	doi = {10.1093/mnras/stz355},
	number = {4},
	urldate = {2023-08-22},
	journal = {Monthly Notices of the Royal Astronomical Society},
	author = {Avdellidou, C and Vaubaillon, J},
	month = apr,
	year = {2019},
	pages = {5212--5222},
}

@article{madiedoLargeLunarImpact2014,
	title = {A large lunar impact blast on 2013 {September} 11},
	volume = {439},
	issn = {0035-8711},
	url = {https://doi.org/10.1093/mnras/stu083},
	doi = {10.1093/mnras/stu083},
	number = {3},
	urldate = {2023-07-24},
	journal = {Monthly Notices of the Royal Astronomical Society},
	author = {Madiedo, José M. and Ortiz, José L. and Morales, Nicolás and Cabrera-Caño, Jesús},
	month = apr,
	year = {2014},
	pages = {2364--2369},
}

@article{bouleyPowerDurationImpact2012,
	title = {Power and duration of impact flashes on the {Moon}: {Implication} for the cause of radiation},
	volume = {218},
	issn = {0019-1035},
	shorttitle = {Power and duration of impact flashes on the {Moon}},
	url = {https://www.sciencedirect.com/science/article/pii/S0019103511004556},
	doi = {10.1016/j.icarus.2011.11.028},
	language = {en},
	number = {1},
	urldate = {2023-07-18},
	journal = {Icarus},
	author = {Bouley, S. and Baratoux, D. and Vaubaillon, J. and Mocquet, A. and Le Feuvre, M. and Colas, F. and Benkhaldoun, Z. and Daassou, A. and Sabil, M. and Lognonné, P.},
	month = mar,
	year = {2012},
	pages = {115--124},
}
\bibliographystyle{aasjournalv7}

\end{document}